\documentclass[twocolumn,prb,showpacs,superscriptaddress,floatfix]{revtex4}

\usepackage{graphicx}
\usepackage{bm}
\usepackage{epstopdf}
\usepackage{amsmath,amsfonts,paralist}

 \let\b=\beta     
     \let\l=\lambda
                
 \let\t=\tau    
   
   \let\Th=\Theta\let\L=\Lambda

\def\PP{{\cal P}}

\def\DD{{\cal D}}\def\AA{{\cal A}}

\def\ul{\underline}

\def\ol{\overline}

\def\to{\rightarrow}

\newcommand{\beq}{\begin{equation}}
\newcommand{\eeq}{\end{equation}}

\newcommand{\Tr}{\text{Tr}}

\begin{document}

\title{The Ising $M$-$p$-spin mean-field model for the structural glass:\\  continuous vs. discontinuous transition}


\author{F.  Caltagirone$^1$, U. Ferrari$^1$, L. Leuzzi$^{1,2}$,
G. Parisi$^{1,2,3}$ and T. Rizzo$^1$} \affiliation{$^1$ Dip. Fisica,
Universit\`a "Sapienza", Piazzale A. Moro 2, I-00185, Rome, Italy \\
$^2$ IPCF-CNR, UOS Rome, Universit\`a "Sapienza", PIazzale A. Moro 2,
I-00185, Rome, Italy \\ $^3$ INFN, Piazzale A. Moro 2, 00185, Rome,
Italy}

\begin{abstract}
The critical behavior of a family of fully connected mean-field models
with quenched disorder, the $M-p$ Ising spin glass, is analyzed,
displaying a crossover between a continuous and a random first order
phase transition as a control parameter is tuned. Due to its
microscopic properties the model is straightforwardly extendable to
finite dimensions in any geometry.
\end{abstract}
\date{\today}

 \maketitle

\section{Introduction }
Since the work of Kirkpatrick, Thirumalai and Wolynes
\cite{Kirkpatrick87a,Kirkpatrick87b,Kirkpatrick87c,
Kirkpatrick88a,Thirumalai88,Kirkpatrick89} a certain set of mean-field
spin-glass models have been shown to own the salient properties of the
behavior of structural glasses. In particular, these models display
dynamic equations that are equivalent to those predicted by the Mode
Coupling Theory (MCT)\cite{Goetze84,Goetze92, Bouchaud96} above the
so-called mode coupling  temperature $T_{\rm mc}$ where
ergodicity breaking occurs in that theory.  Moreover, two kinds of
transition are predicted: a dynamic one at $T_{d}=T_{\rm mc}$ and a
thermodynamic phase transition at a lower $T$, often referred to as
Kauzmann transition.  Mean-field models exhibiting structural glass
features are characterized by multi-body microscopic interactions and
their thermodynamics is self-consistently described by implementing a
discontinuous Replica Symmetry Breaking (RSB) Ansatz (usually one
step: 1RSB).
\\
\indent
The dynamic transition is due to the presence of a large number of
 metastable excited glassy states, represented as local minima in the
 free energy landscape in the configurations space, growing
 exponentially with the size $N$ of the system.  In the mean-field
 approximation, barriers between minima grow with the size, so that,
 in the thermodynamic limit, the relaxing dynamics to equilibrium of
 the system at $T\leq T_d$ remains stuck forever inside the first
 ``meta''-stable state where it ends up in.  In real glassy systems,
 however, there is a slow dynamics occurring through activated
 processes and this dynamic arrest is an artefact due to the
 mean-field approximation. In finite dimensions the glass transition
 occurs because at some glass temperaure $T_g$ the time-scales of
 observation are shorter than the characteristic time-scales of the
 slowest structural $\alpha$ processes taking place in the
 glass-former sample.  Metastable states really have a finite
 time-life, even though (much) longer than the experimental time of
 observation.  The effect of activated processes starting from
 spin-glass 1RSB mean-field models has been analyzed, e.g., by working
 at finite $N$ in the fully connected Random Orthogonal Model (ROM)
 \cite{Crisanti00a,Crisanti00b, Crisanti00c,Rao03b} and finding a
 glass behavior, similar to the one observed in computer glasses, cf.,
 e.g., Refs. \cite{Sciortino05,LeuzziBook}.
\\
\indent
As we mentioned, another property occurring in the
glass-like mean-field models (see also
Refs. \cite{Gross84,Gross85,Crisanti92}) is a thermodynamic transition
between the supercooled liquid (below $T_d$) and a thermodynamically
stable glass.  This occurs with a jump in the order parameter, but
without discontinuity in the internal energy (no latent heat is
exchanged). This mixture of first order and continuous phase
transition in presence of disorder has been termed random first order (RFOT).
\cite{Kirkpatrick89}
\\
\indent
One of the most accredited theories, the
Adam-Gibbs-Di Marzio entropic theory \cite{Gibbs58,Adam65} predicts
the existence of a thermodynamic transition to an {\em ideal glass}
phase, the so-called Kauzmann transition. The Kauzmann temperature is
generally associated with the asymptote of the Vogel-Fulcher law
\cite{Vogel21, Fulcher25} of the relaxation time and, thus, related to
the transition one might have in a infinitely slow cooling of a
never-crystalizing glass-former.  Because of the impossibility of
experimental measurements of glass relaxation in those conditions, the
very existence of the Kauzmann point and the nature of that transition
is still a matter of debate.
\\
\indent
Attempts to follow the properties envisaged in mean-field models in
realistic systems have faced the problem of finding a proper way of
embedding the model microscopic features into a given finite
dimensional geometry (e.g., 3D cubic lattice) without altering the
discontinuous nature of the transition.  So that one can actually
falsify the hyphothesis of RFOT in finite dimensional systems.
Indeed, 
in Ref. [\onlinecite{Parisi99}] a generalization of the $p=3$-spin model with
$M=2$ Ising spins on each site was numerically studied on a $D=4$
hypercubic lattice finding evidence for a continuous phase transition.
The same continuous behavior was recently found, already in the
mean-field regime, in the same $p=3$, $M=2$ model in a $D=1$ chain
 on a ``Levy lattice''.\cite{Larson10}
\\
\indent
Starting from this observation, that the RFOT becomes
continuous in finite $D$,
the work of Moore, Drossel and Yeo \cite{Moore02,Yeo06,Moore06}
 shows that this is equivalent to the
critical behavior of the Edwards-Anderson model in a field, where the
transition line is called de Almeida-Thouless (dAT) line.  Applying
droplet theory (that rules out the existence of a dAT line outside the
limit of validity of mean-field theory) it is, thus, inferred that no
thermodynamic random first order transition can occur in real structural glasses.
The issue of the existence of a dAT line in finite dimensional
spin-glasses will not be addressed here.  For recent bibliography on
that subject see
Refs. [\onlinecite{Jonsson07,Jorg08,Katzgraber09,Leuzzi09,Tabata10,Leuzzi10}]
and references therein.
\\ \indent In the present work we will focus on deriving a mean-field
class of models, to whom Ising $p$-spins \cite{Gross84,Gardner85}
belong, whose critical behavior shifts from continuous to
discontinuous in a controlled way.  The aim is to clarify why the
finite dimension extensions of mean-field glasses studied so far do
not display RFOT and to devise  mean-field models whose
discontinuous critical nature can be conserved also beyond the limit
of validity of the mean-field approximation.  The model consists of $N$
sites, each one containing $M$ spins interacting with spins on other
sites in $p$-uples.  We will see how, changing $p$ and the number $M$
of spins living on a single site, it is possible to move from systems
displaying a second order phase transition to systems displaying a
random first order transition, that is, yielding both a dynamic and a
Kauzmann-like phase transition.  The finite dimension counterpart of
the model under probe can be easily achieved since the $p$-spin
interaction is always exchanged between two sites, e.g., nearest
neighbors on a $d$-dimensional (hyper)cubic lattice. 
\\
\indent
We mention that
moving from mean-field to finite dimensions, also standard Ising $p$-spin
and Potts models might conserve the random first order nature of the transition and keep reproducing
basic features of structural glasses. Even though it is not straightforward
to conceive a short-range finite dimensional Ising $p$-spin, the Potts model can
be easily defined on a hyper-cubic lattice. Nevertheless, no numerical
evidence has been collected so far for a discontinuous RFOT in disordered Potts models with number of states $p_{\rm Potts}=5,6,10$ \cite{Brangian03,Lee06,Banos10} and,
actually, we found no argument to infer that in the finite dimensional lattice case the $p_{\rm Potts} \to
\infty$ limit can be kept under control. 
\\
\indent 
 On the other hand, the model
considered in the present work has the advantage to reduce to an exact
mean-field model for the RFOT
as $M\to\infty$ even in finite dimension (and finite size), for any values of $p$.
Moreover, we can work out a sufficient criterion to determine the smallest value of $M$ above which continuous transitions cannot occur.
\\
\indent
The manuscript is organized as follows:
in Sec. \ref{s:model}  we will study
the statistical mechanics of the model; in Sec. \ref{s:largeM} we show that the 
large $M$ limit  corresponds to standard $p$-spin and in Sec. \ref{s:crit}, expanding near
criticality, we build the corresponding field-theory, compute the
coupling constants and study the relevance of terms competing for
continuous/discontinuous transition.
In Sec. \ref{s:conclusions} we present our conclusions.

\section{The Model}
\label{s:model}
The model consists on $N$ sites, each one hosting a set of $M$ spins.
Two sites interact through a $p$-body interaction involving 
spins belonging to the two sets of $M$ spins.
The Hamiltonian reads
\begin{equation}
{\cal H}= - \sum_{\langle x, y \rangle} \sum_{g(x,y)} \, J_g \prod_{\mu \in
g}\, s_{\mu}
\end{equation}
where $<x,y>$ indicates the sum over all couples of sites and $g(x,y)$
are all the possible $p$-uplets among the $2M$ spins, with an
exception if $p\leq M$: those $p$-uplets completely pertaining to a single
site are excluded. This choice actually {\em defines} our model when
$p\leq M$, as we will discuss in the following.

The disordered interactions are Gaussian i.i.d. variables, with distribution:
\beq
P(J_g)=\frac{1}{\sqrt{2\pi \sigma_J^2}} e^{-\frac{J_g^2}{2\sigma_J^2}}
\eeq
where, to provide the right thermodynamic convergence of the free energy,
the variance scales like 
\beq
\sigma_J^2=\frac{1}{N M^{p-1}}
\eeq

\subsection{Free energy and order parameters}
Replicating $n$ times the system we compute the average over quenched
disorder of the replicated partition function:
\begin{eqnarray}
\ol{Z^n}&=&\int\, \prod_{\langle x, y \rangle}^{1,N}\, \prod_{g(x,y)}
P(J_g) \, dJ_g 
\\
\nonumber
&&\, \times Tr_{[s]} \, \exp \Bigl[\beta\, \sum_{a=1}^n \sum_{\langle
x, y \rangle}^{1,N} \sum_{g(x,y)} J_g \prod_{\mu \in g}\, s^a_{\mu}
\Bigr]
\end{eqnarray}
yielding
\beq \ol{Z^n} = Tr_{[s]}\, \exp \Biggl[ \frac{\beta^2}{4NM^{p-1}}
\sum_{x\neq y }^{1,N} \sum_{g(x,y)} \sum_{a,b}^{1,n}\prod_{\mu \in g}
s^a_{\mu}s^b_{\mu} \Biggr] 
\eeq

Explicitly separating those spins belonging to site $x$ from those on
site $y$ one can obtain a general expression for the partition
function valid both for $p > M$ and $p \leq M$:
\begin{eqnarray}
&&\hspace*{-.3cm}\ol{Z^n}  =\Tr_{\{s(x), s(y)\}}\, \exp \Biggl[
\frac{\beta^2}{4NM^{p-1}}
\\
\nonumber
&&\hspace*{-.3cm}\times
  \sum_{a,b}\,
\sum_{x\neq y } \sum_{k}
\sum_{ i_1< \dots < i_k }\hspace{-.2cm}
s^a_{i_1}(x)s^b_{i_1}(x)\dots
s^a_{i_k}(x)s^b_{i_k}(x)
\label{f:Zn_xy_k}
\\
\nonumber
&&\qquad\quad \,\,\,
\sum_{i_{k+1}< \dots < i_p}\hspace{-.2cm}
s^a_{i_{k+1}}(y)s^b_{i_{k+1}}(y)\dots
s^a_{i_p}(y)s^b_{i_p}(y) 
\Biggr]
\end{eqnarray}
 For $p > M$ the sum over $k$ runs from $p-M$ to $M$; in the case
$p\leq M$ the sum over $k$ runs from $1$ to $p-1$. In principle, it might
be possible to include an extra term due to self-interaction: $p$ out
of $M$ spins interact on a single site (``a single site standard
$p$-spin'').  As already mentioned, in the present work we will
consider a model {\em without} site self-interaction.
We now introduce a set of multi-overlaps between $k$ spins on the same
site $x$ in two replicas:

\begin{eqnarray}
&&\hspace*{-.2cm} Q_{ab}^{(k)}\equiv \frac{1}{NM^k} 
 \sum_{x=1}^N 
  \sum_{i_1<\ldots < i_k}
 \hspace{-.2cm} s^a_{i_1}(x)s^b_{i_1}(x).. s^a_{i_k}(x)s^b_{i_k}(x)
 \nonumber
 \\
  \end{eqnarray}
with 
\begin{eqnarray}
\nonumber
k=p-M, \ldots, M \qquad &\mbox{if }&  p>M
\\
\nonumber
k=1,\,\,\, \ldots \,\,\,, p-1 \qquad &\mbox{if }&  p\leq M
\end{eqnarray}
 By means of multi-overlaps we can write the replicated partition
function Eq. (\ref{f:Zn_xy_k}) as
\begin{eqnarray}
&&\ol{Z^n}  = e^{NC}\,\int \, D\ul{Q} \, \Tr_{\{s(x), s(y)\}}\, 
\label{f:Zn1}
\\
\nonumber
&&\exp
\Bigl[ \frac{\beta^2 NM}{4} \sum_{k }\sum_{a\neq b}\, Q_{ab}^{(k)}
Q_{ab}^{(p-k)}\Bigr] \, \times 
\\
\nonumber
 && \times \prod_k \prod_{a< b}
\delta \Biggl(NM Q_{ab}^{(k)} 
\\
\nonumber
&&- \frac{1}{M^{k-1}}\sum_x 
\sum_{i_1<\dots < i_k}s^a_{i_1}(x)s^b_{i_1}(x)\dots
s^a_{i_k}(x)s^b_{i_k}(x)\Biggr)
\end{eqnarray}
where the parameter $C$, proportional to minus the paramagnetic free energy, reads
\begin{eqnarray}
&&\frac{C}{n}=\frac{\beta^2}{4M^{p-1}}\, \Biggl[ \sum_k \, \binom{M}{k} \binom{M}{p-k} \Biggr]
\\
\nonumber
&&=\frac{\beta^2}{4M^{p-1}}\, \left\{\begin{array}{l} 
  \!\! \frac{1}{\Gamma(1 + p)}
 \!\!
 \left[ 
 \frac{\Gamma(1 + 2 M)}{\Gamma(1 + 2 M - p)}
-\frac{2 \Gamma(1 + M)}{\Gamma(1 + M - p)} \right] 
 \\
 \hspace*{3.8cm} p\leq M 
 \\
 \vspace*{-.1cm}
 \\
 \!\!\frac{1}{\Gamma(1 + 2 M - p)}
\!\!\left[
 \frac{2\Gamma(1 + M)}{\Gamma(1 - M + p)}
-\frac{\Gamma(1 + 2 M)}{\Gamma(1 + p)} 
\right] 
\\
\hspace*{3.8cm} p>M 
 \end{array}
 \right.
 \end{eqnarray}

Introducing the integral representation for the delta functions in
Eq. (\ref{f:Zn1}) one obtains:
\beq
\begin{split}
\overline{Z^n} & = e^{NC}\,\int \,  D\ul{Q}\, D\ul{\Lambda} \, \exp[-NG(\ul{Q},\ul{\Lambda})]
 \end{split}
\eeq
\begin{eqnarray} 
G(\ul{Q},\ul{\Lambda})&=&-\frac{\beta^2 M}{4} \sum_{k}\sum_{a\neq
b}\, Q_{ab}^{(k)} Q_{ab}^{(p-k)} 
\\
\nonumber
&&+ \frac{M}{2}\sum_{k}\sum_{a\neq b}
\, \,\Lambda_{ab}^{(k)}Q_{ab}^{(k)} - \log Z(\ul{\Lambda}) 
\end{eqnarray}

\begin{eqnarray}
 &&Z(\ul{\Lambda}) =Tr_{[s]} e^{S(\ul{\L})}
  \\
 \nonumber
&& S(\ul{\L})=\frac{1}{2}\sum_k \sum_{a\neq b}\,
\frac{\Lambda_{ab}^{(k)}}{M^{k-1}}\sum_{i_1<\dots< i_k}\,
s^a_{i_1}s^b_{i_1}\dots s^a_{i_k}s^b_{i_k} 
\end{eqnarray}

\beq
\begin{split}
& D\ul{Q}=\prod_{k}\prod_{a<b}\,\, dQ_{ab}^{(k)}\\
& D\ul{\L}=\prod_{k}\prod_{a<b}\,\, d\L_{ab}^{(k)}
\end{split}
\eeq
The stationarity equations in $\L$ and $Q$ are
\begin{eqnarray}
 Q_{ab}^{(k)}&=& \frac{1}{Z(\ul{\Lambda})}\Tr_{[s^a]} \frac{1}{M^k}
\label{f:Qsta}
\\
&&\times
\sum_{i_1<\dots< i_k} s^a_{i_1}s^b_{i_1}\dots s^a_{i_k}s^b_{i_k}
e^{S(\ul{\Lambda})} 
\nonumber
\\
\Lambda_{ab}^{(k)}&=& \beta^2\, Q_{ab}^{(p-k)}
\end{eqnarray}
Substituting the saddle point value for $\L$ in the effective action
we obtain
\begin{eqnarray}
 &&\hspace*{-.3cm}G(\ul{Q})=\frac{\beta^2 M}{4} \sum_{k}\sum_{a\neq b}\,
Q_{ab}^{(k)} Q_{ab}^{(p-k)} 
\label{f:actionQ}
\\
\nonumber
&&\qquad\qquad\qquad- \log \Tr_{[s^a]} e^{S(\ul{Q})}
\\
\nonumber
&&\hspace*{-.3cm}S(\ul{Q})=
\frac{\beta^2}{2} 
\sum_k \sum_{a \neq b}
\frac{Q_{ab}^{(p-k)}}{M^{k-1}}\sum_{i_1<\dots< i_k}
s^a_{i_1}s^b_{i_1}\dots s^a_{i_k}s^b_{i_k}
\end{eqnarray}

The physical meaning of the overlap matrix at saddle point value is
the usual one and, more precisely
\begin{eqnarray}
&&Q_{ab}^{(k)}=\frac{1}{N M^k} \sum_{x=1}^N \sum_{i_1<i_2<\dots <i_k}
 \ol{\langle s_{i_1}(x) \dots s_{i_k}(x) \rangle ^2}
\nonumber
\\
&&\qquad=\lim_{n\to 0} \frac{2}{n(n-1)} \sum_{a<b} Q^{(k)}_{ab} \big\arrowvert_{SP}
\end{eqnarray}

\section{Large $M$ limit: standard $p$-spin}
\label{s:largeM}
For large $M$, neglecting diagonal terms in the sum over $i_1,\ldots,i_k$,
in Eq. (\ref{f:actionQ}), the $\log\Tr$ term  can be rewritten as
\begin{equation}
S(\ul{Q})=M\frac{\beta^2}{2} 
\sum_{k=1 }^{p-1}\sum_{a \neq b}
Q_{ab}^{(p-k)}\frac{1}{k!}\left(\frac{1}{M}\sum_{i=1}^M 
s^a_{i}s^b_{i}\right)^k
\end{equation}

Performing the saddle point for large $M$, rather than $N$, and introducing the auxiliary parameter
\begin{equation}
q_{ab}\equiv  \frac{1}{M}\sum_{i=1}^M 
s^a_{i}s^b_{i}
\end{equation}
we obtain, for the free energy Eq. (\ref{f:actionQ})
\begin{eqnarray}
&&G(\ul{Q})=M\Biggl[ \frac{\beta^2 }{4} \sum_{k=1}^{p-1}\sum_{a\neq b}\,
Q_{ab}^{(k)} Q_{ab}^{(p-k)} 
\label{f:free_largeM}
\\
\nonumber
&& -\frac{\beta^2}{2} 
\sum_{k=1}^{p-1} \frac{1}{k!}\sum_{a \neq b} Q_{ab}^{(p-k)} q_{ab}^k
+\lambda_{ab}q_{ab}
\\
\nonumber
&&
- \log \Tr_{[s^a]} \exp \Biggl\{\sum_{a\neq b}\lambda_{ab}s^as^b\Biggr\}
\Biggr]
\nonumber
\end{eqnarray}
The saddle point self-consistency equation w.r.t. $Q^{(p-k)}$
yields 
\begin{equation}
Q_{ab}^{(k)}=\frac{1}{k!}q_{ab}^k
\label{f:sp_largeM}
\end{equation}
Substituting Eq. (\ref{f:sp_largeM}) in Eq. (\ref{f:free_largeM}), we obtain the expression
\begin{eqnarray}
&&G(\ul{q},\ul{\lambda})/M=-\frac{\beta^2 }{4} \sum_{a\neq b}\sum_{k=1}^{p-1}\, \frac{1}{k!(p-k)!}q_{ab}^p
+\lambda_{ab}q_{ab}
\nonumber
\\
&&
- \log \Tr_{[s^a]} \exp\Biggl\{ \sum_{a\neq b}\lambda_{ab}s^as^b\Biggr\}
\end{eqnarray}
\noindent
that is, the standard formal free energy of the fully connected Ising $p$-spin model:
\begin{eqnarray}
&&G(\ul{q},\ul{\lambda})/M=-\frac{\beta^2 }{4} \sum_{a\neq b}\frac{2^p-2}{p!}q_{ab}^p
+\lambda_{ab}q_{ab}
\nonumber
\\
&&
- \log \Tr_{[s^a]} \exp\Biggl\{ \sum_{a\neq b}\lambda_{ab}s^as^b\Biggr\}
\label{f:pspinG}
\end{eqnarray}
with 
\begin{eqnarray}
q_{ab}&=&\langle s^a s^b\rangle
\\
\lambda_{ab}&=&\frac{p\beta^2}{2} q_{ab}^{p-1} \  .
\end{eqnarray}

\section{Analysis of the Critical point}
\label{s:crit}
Our aim is to find the transition point and to study its thermodynamic
nature as $M$ and $p$ are changed. In particular, we will verify that,
at given $p$ (vice-versa $M$) there are threshold values of $M$
(resp. $p$) beyond which the transition switches from continuous to
discontinuous.

First, to identify the critical point we expand the stationarity
 equation (\ref{f:Qsta}) to first order in $Q_{ab}^{(k)}$, obtaining:
\beq
Q_{ab}^{(k)}=\frac{\beta^2}{M^k}\frac{Q_{ab}^{(p-k)}}{M^{k-1}}\binom{M}{k}
\eeq
There are ``multi''- critical temperatures for the  ``multi'' -
overlaps, whose expressions read
\beq
\beta_c(k)=\frac{M^{\frac{p-1}{2}}}{\binom{M}{k}^{\frac{1}{4}}\binom{M}{p-k}^{\frac{1}{4}}}
\label{f:betac}\eeq
The largest critical temperature is obtained for $k=p/2$ if $p$ is
even, and for $k=(p+1)/2\, , \, (p-1)/2 $ if $p$ is odd.  The overlap
corresponding to the smallest $\beta_c$ (slightly above $\b_c$) is
non-zero and of order $\t \propto (T_c - T)/T_c $, while the others are at least of order
$\t^2$.

Proceeding to the second order expansion of Eq. (\ref{f:Qsta}) we have
\begin{eqnarray}
&&\hspace*{-.2cm}Q_{ab}^{(k)} =  \frac{\beta^2}{M^k}\frac{Q_{ab}^{(p-k)}}{M^{k-1}}\binom{M}{k}
\label{f:Qsta2}
\\
\nonumber
&&+\Tr_{[s^a]} \frac{1}{M^k} \sum_{i_1<\dots< i_k}
s^a_{i_1}s^b_{i_1}\dots s^a_{i_k}s^b_{i_k} 
 \\
  &&
  \nonumber
  \times
\frac{\b^4}{4\times 2!}\sum_{l,m} \sum_{c\neq d, e\neq f}\,
\frac{Q_{cd}^{(p-l)}}{M^{l-1}}\frac{Q_{ef}^{(p-m)}}{M^{m-1}} 
\\
\nonumber
&&\times \sum_{j_1<\dots< j_l}\, \sum_{t_1<\dots< t_m}
s^c_{j_1}s^d_{j_1}\dots s^c_{j_l}s^d_{j_l}\,
s^e_{t_1}s^f_{t_1}\dots s^e_{t_m}s^f_{t_m}
\end{eqnarray}
we will focus only on the equations for the overlaps corresponding to
the largest critical temperature, cf. Eq. (\ref{f:betac}), that is,
on the terms of the type 
\beq
\nonumber
Q_{ab}^{(p/2)}Q_{ab}^{(p/2)}, \quad \text{for even }
p
\eeq
 else 
 \beq
 \nonumber
 Q_{ab}^{(\frac{p \pm 1}{2})}Q_{ab}^{(\frac{p \pm 1}{2})}, \quad
\text{for odd } p.
\eeq
More specifically, we are interested in the terms of the series at the
r.h.s of Eq. (\ref{f:Qsta2}) with $k=l=m=p/2$, if $p$ is even,
or with  $k,l,m=\frac{p \pm
1}{2}$, if $p$ is odd.

It is interesting to notice that we would have the same physics considering a model in which 
$p/2$-uples on each site interact with $p/2$ on another site ($p$ even) or 
$(p+1)/2$-uples on a site interact with $(p- 1)/2$ on another site ($p$ odd).

In Eq. (\ref{f:Qsta2})
each spin in each replica has to be matched with another one in
 another replica in order to get a non-zero result from the trace.  At
 second order we are, thus, left with only two kinds of possible matching,
 yielding terms:
\beq \nonumber \sum_c
Q_{ac}^{(\times)}Q_{cb}^{(\times)}=  \left(Q^{(\times)}\right)^2_{ab} \qquad \text{and}
\qquad \left( Q_{ab}^{(\times)} \right)^2.  \eeq 
We will see how, depending on the parity of $p$ even the multiplicity of such
terms will change, leading to different expressions of their coefficients as functions of $p$ and $M$.

Using the above results, Eq. (\ref{f:actionQ}), approximated to the
second order in $Q$, can then be written as
\beq
 G(\ul{Q}) = \frac{\tau}{2} \sum_{a,b} Q_{ab}^2 
+\frac{w_1}{6} \Tr Q^3 
+\frac{w_2}{6} \sum_{a,b}Q_{ab}^3
 \label{f:actionQ3}
 \eeq 
where $Q_{ab}$ stays for $Q_{ab}^{(\times)}$.

As already noticed by Gross, Kanter and Sompolinsky \cite{Gross85} in
the Potts model (threshold was $p_{\rm Potts}=4$ colors) and in
Ref. [\onlinecite{Ferrero96}], it can be shown (see
App. \ref{thre}) that if the ratio between coupling constants on the
nonlinear terms is larger than one the phase transition cannot be
continuous.  We will now proceed to the computation of the coupling
constants for the $M$-$p$ Ising spin model. Since, as already
mentioned, the computation of the third order coefficients will yield
different functional expressions depending on the values of $p$, we
have to distinguish between four cases:
\beq p=\left\{\begin{array}{c r} 4a &
\textbf{A.}  \\ 4a-2& \textbf{B.} \\ 4a-1 &\textbf{C.}  \\ 4a-3
&\textbf{D.}
\end{array} 
\right.
\quad  a\in \mathbb{N}^+
\eeq 
and we will analyze them separately.

\subsection{Even $p$ and $p/2$, $p=4a$}
The only surviving term in the sum over $l$ and $m$
in the r.h.s. of Eq. (\ref{f:Qsta2}) is for $l=m=k=p/2$. 
The trace term turns out to be
\begin{eqnarray}
&&w_1\sum_{c=1}^n Q_{ac}^{(p/2)} Q_{cb}^{(p/2)}
\\
\nonumber
&&=\frac{\beta^4}{M^{3 p/2-2}}
\binom{M}{ p/2}
\sum_{c=1}^n Q_{ac}^{(p/2)} Q_{cb}^{(p/2)}
\end{eqnarray}
The squared term is
 \begin{eqnarray} 
 &&w_2 \left( Q_{ab}^{(p/2)} \right)^2
 \\
 \nonumber
 &&=\frac{1}{2
 M^{\frac{3}{2}p-2}} \binom{M}{p/2} \binom{p/2}{p/4}
 \binom{M-p/2}{p/4}\left( Q_{ab}^{(p/2)} \right)^2 \end{eqnarray}
 and the ratio
 \beq
 \frac{w_2}{w_1}=\frac{1}{2}\binom{p/2}{p/4}\binom{M-p/2}{p/4}
 \label{f:ratio_pp4}
 \eeq

\subsection{Even $p$, odd $p/2$, $p=4a-2$}
In the r.h.s. of Eq. (\ref{f:actionQ3}) only the coefficient in front of
 the $\Tr \, Q^3_{ab} $ term survives, whereas $w_2=0$ always. The
 ratio is
 \beq
 \frac{w_2}{w_1}=0
\label{f:ratio_ppn4} \eeq
According to the small $Q$ expansion, Eqs. (\ref{f:Qsta2}), (\ref{f:actionQ3}), when
 $p$ is even and $p/2$ is odd the transition at the largest critical temperature
$1/\beta_c(p/2)$, cf. Eq. (\ref{f:betac}), turns out can to be continuous, no
matter how many spins $M$ stay on each site.
This might appear at contrast with the large $M$ limit leading to  Eq. (\ref{f:pspinG}) that is equivalent to
an Ising $p$-spin mean-field model for any $p>2$. There, however, no perturbative expansion was carried out, while Eq. (\ref{f:ratio_ppn4}) is the outcome of an expansion for small overlap values that cannot help in identifying discontinuous transitions with large jumps in $Q$. Indeed, {\em the condition expressed by Eq. (\ref{f:cond}) is sufficient but not necessary to rule out a continuous transition}.

 \subsection{Odd $p$, even $(p+1)/2$, $p=4a-1$}

When $p$ is odd we have to deal with two relevant overlaps
$Q^{(\frac{p-1}{2})}$ and $Q^{(\frac{p+1}{2})}$ and one critical
temperature. In order to determine the coupling constants of the cubic
terms one, thus, has to diagonalize a $2\times 2$ matrix.  In
App. \ref{app} we report the details of the computation leading to:
 \beq
  w_1(p,M)
=\frac{\sqrt{2M-p+1}+\sqrt{p+1}}{4M^{p-3/2}\sqrt{p+1}}
\sqrt{\binom{M}{\frac{p-1}{2}}}
\eeq
%
for the coefficient of the cubic trace term in the action,
Eq. (\ref{f:actionQ3}).  The expression for the coupling constant
depends, further, on $(p+1)/2$ being even or odd.  For $p=4a-1$ we
obtain
   \begin{eqnarray}
 w_2(p,M) &=&\frac{1}{8M^{p-3/2}}
 \frac{5M-3p+2}{2M-p+1}
 \\
 \label{f:w1_oddp}
 &&
 \nonumber
\times \sqrt{
 \binom{M}{\frac{p-1}{2}
 }}
  \binom{\frac{p+1}{2}}{\frac{p+1}{4}}
 \binom{M-\frac{p-1}{2}}{\frac{p+1}{4}}
\end{eqnarray}
and, eventually, the ratio is:
\begin{eqnarray}
\frac{w_2}{w_1}&=&\frac{2 + 5 M - 3 p}{2 + 4 M - 2 p} 
\frac{\sqrt{p+1}}{\sqrt{p+1} + \sqrt{2M-p+1}}
\nonumber\\
&&
\label{f:ratio_4a-1}
\times\binom{ M - \frac{p-1}{2}}{ \frac{1 + p}{4}} 
\binom{\frac{ 1 + p}{2}}{ \frac{1 + p}{4}}
\end{eqnarray}

\subsection{Odd $p$, even $(p-1)/2$, $p=4a-3$}
In this last case the coupling of the trace cubic term is still given
 by Eq. (\ref{f:w1_oddp}) and the second nonlinear coupling constant
 is expressed as:
 \begin{eqnarray}
 \nonumber
 w_2(p,M)& = &\frac{1}{4M^{p-3/2}}
 \frac{6 - p - 5 p^2 +9 M +7 p M}{(p+3)(2M-p+1)}
\\
&&
\times\sqrt{\binom{M}{\frac{1+p}{2}}}
  \binom{M-\frac{p-1}{2}}{\frac{p-1}{4}}
 \binom{\frac{p-1}{2}}{\frac{p-1}{4}}
\end{eqnarray}
yielding the ratio:
\begin{eqnarray}
\frac{w_2}{w_1}&=&
\frac{6 - p - 5 p^2 +9 M + 7 p M}{(p+3)(2M-p+1)}
\label{f:ratio_4a-3}
\\
\nonumber
&&\!\!\!
\times \frac{\sqrt{2M-p+1}}
{\sqrt{p+1} + \sqrt{2M-p+1}} \binom{M-\frac{p-1}{2}}{\frac{p-1}{4}}
 \binom{\frac{p-1}{2}}{\frac{p-1}{4}}
\end{eqnarray}

We now have a complete description of the critical behavior of the
$M$-$p$ system.  Already at the mean-field level, in order to have a
discontinuous transition a $p>2$ interaction between spins in not
enough.  We find that for each given $p$ one needs a minimal number of
spins $M_{\rm disc}$ on each site in order to have a random first
order phase transition, corresponding to the lowest integer $M$ for
which $w_2/w_1>1$; cf.  Eq. (\ref{f:ratio_pp4}), (\ref{f:ratio_4a-1}),
or (\ref{f:ratio_4a-3}) depending on the parity of $p$ and $(p+1)/2$.

\begin{figure}
\includegraphics[width=.99\columnwidth]{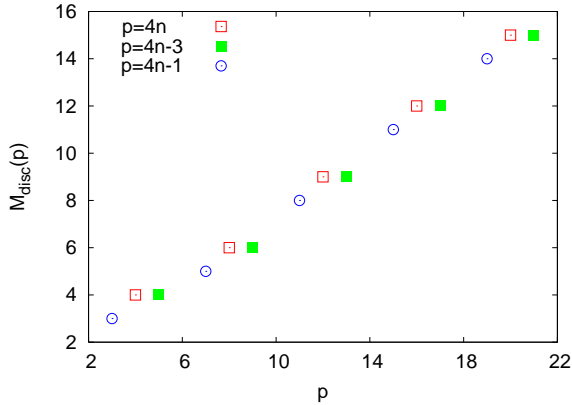}
\caption{Lowest integer values of $M$ at given $p$, for which a
discontinuous transition is certainly expected (sufficient condition to have RFOT is $M \geq M_{\rm disc})$.}
\label{fig:Mp}
\end{figure}

In table Tab. \ref{tab:Mp} we report some values of the ratios for
systems with small $p$ and $M$.  In Fig. \ref{fig:Mp} we plot the
$M_{\rm disc}(p)$ behavior.
\begin{table}[!t]
\begin{tabular}{| l | c | c |}\hline \hline
$p $     &    $M$  &    $w_2/w_1$  \\ \hline
3   &    2 & $3\left(1-1/\sqrt{2}\right)=0.87868$ \\
3 & 3 & 2 \\
4 & 2 & 0\\
4 & 3  & 1\\
4 & 4  & 2\\
5& 3  & $(\sqrt{3}-1)/2=0.366025$\\
5& 4 & $13\left(\sqrt{3/2}-1\right)=2.92168$\\
6 & {\mbox{any}} & 0
\\ \hline
\end{tabular}
\caption{Ratio values for small  $p$ and $M$ around the threshold $1$}
\label{tab:Mp}
\end{table}

\section{Conclusions}
\label{s:conclusions}
In the present work we have performed an analytic computation of the
critical behavior of a mean-field $p$-spin model that can display both
a random first order and a continuous phase transition tuning the
number of spins present on each site.  The effective action is
represented by Eq. (\ref{f:actionQ3}) where the first two relevant
terms are third order in the overlap parameter $Q$.  The ratio of their
coupling constants has been computed, thus providing the threshold
values of $M_{\rm disc}(p)$ separating models with continuous from models with
discontinuous critical behavior. If $M\geq M_{\rm disc}$ this is sufficient to guarantee the discontinuity of the transition, though it is not necessary. A RFOT can also occur at lower values of $M$ that cannot be identified with our probe based on the perturbative expansion for small Q, cf. Eq. (\ref{f:actionQ3}). 
\\ \indent The particular case studied in
Ref. [\onlinecite{Parisi99}], ($M=2$, $p=3$), yielded numerical
evidence for a continuous phase transition in dimension four. This is
consistent with the value of the $w_2/w_1=3[1- 1/\sqrt{2}]
=0.87868$ as computed in the mean-field theory, cf. Tab. \ref{tab:Mp}
and Ref. [\onlinecite{Yeo}].  The same applies to the model recently studied in
Ref. [\onlinecite{Larson10}], a one dimensional $(M,p)=(2,3)$ model on
a Levy lattice \cite{Leuzzi08}.
\\ \indent The continuous-discontinuous cross-over is like the one
found in Potts \cite{Gross85} varying the number of colors, and in the
spherical $p$-spin varying an external magnetic field
\cite{Crisanti92}.  The advantage of the present model is that it can
be easily represented in finite dimensions on lattices of given
geometry, e.g., on a cubic lattice with short-range interactions, and that it always displays a RFOT in the $M \rightarrow \infty$
limit for all lattices, both finite dimensional and mean-field like.

\acknowledgements
We thank Joonhyun Yeo and Mike A. Moore for interesting discussions and for sharing with us their unpublished results.

\appendix
\section{Threshold value for $w_2 / w_1 $}
\label{thre}
 Starting from the self-consistency equation
for small $Q$'s (implying a continuous transition in $Q$),
cf. Eq. (\ref{f:actionQ3}) where also the quartic term is considered,
\beq
\t Q_{ab}+w_1 (Q^2)_{ab}+w_2 Q_{ab}^{2} + y Q_{ab}^{3}=0
\label{f:spQ}
\eeq
we have, in the RSB Ansatz,  
\begin{eqnarray}
&&\t q(x) - w_1 \Bigl[ 2 \, q(x) \int_0^1 q(s)\, ds 
\label{f:spqx}
\\
\nonumber
&&+ \int_0^x [q(x)-q(s)]^2\, ds \Bigr]
+ w_2 q(x)^2+ y q(x)^3=0
\end{eqnarray}
Deriving once Eq. (\ref{f:spqx}) w.r.t. $x$ one has
\begin{eqnarray}
q'(x)&& \Biggl\{ \t -2 w_1 \left[ \int_x^1 q(s)\, ds+x q(x)\right] 
\\
\nonumber
&&\qquad\qquad
+ 2 w_2 q(x) + 3 y q(x)^2\Biggr\}=0
\end{eqnarray}
If $q'(x)\neq 0$, deriving a second time w.r.t. $x$, one finds
\beq
q'(x) \left[ -w_1 x + w_2 +3 y q(x) \right]=0
\eeq
If $y>0$, then the overlap function around
criticality can be written as:
\beq q(x) =\left\{
\begin{array}{c l}
 0 & x < x_1 \\
 \frac{1}{3y} (w_1 x - w_2) & x_1< x <x_2 \\
 q(1) & x>x_2\\
\end{array} \right.
\eeq
where, for continuity, $x_1=w_2/w_1$ and
\beq
q(1)={\tau \over p}\,{2 \over 1+\sqrt{1-6 y \tau /p^2}}, \quad p\equiv w_1(1-x_1)
\eeq

As a consequence,
 in order to have a continuous transition it must be
 \beq
\frac{w_2}{w_1}\leq 1.
 \label{f:cond}
 \eeq

The argument for the threshold value of $x$ still
works also if $y\leq 0$ in Eq. (\ref{f:spQ}). In that case, rather than a continuous function,
we simply have a 1RSB step function for 
$q(x)=\theta(x-x_1)q$, 
with 
 \begin{eqnarray}
 &&q={\tau \over p}\,{2 \over 1+\sqrt{1-10 y \tau /p^2}} 
 \\
 \nonumber
 && p\equiv {w_1\left(1-\frac{
 w_2}{w_1}\right)}
 \\
 &&
 x_1={w_2 \over w_1}+{3 y q \over w_1}
 \end{eqnarray}

\section{Coupling constants with odd $p$}
\label{app}
When $p$ is odd we have to deal with two relevant overlaps and one
critical temperature.  The
second order equation, Eq. (\ref{f:Qsta2}), has the structure:

\beq
\AA \mathbf{Q}_{ab}= \mathbf{F}(\{ \mathbf Q \})
\label{f:AQ}
\eeq
where $\mathbf{Q}_{ab}=\{Q_{ab}^{(p-M)}, \cdots, Q_{ab}^{(M)}\}$. Diagonalizing $\AA\to \DD_{\AA}=\PP^{-1} \AA \PP$ one obtains:
\beq
\label{f:Adiag}
\PP^{-1} \AA \PP \PP^{-1} \mathbf{Q}_{ab}= \PP^{-1} \mathbf{F}(\{ \mathbf{Q}_{ab} \} )
\eeq
Introducing new variables $\mathbf{\Th}_{ab}$, linear combinations of $\mathbf{Q}_{ab}$, the above expression can be rewritten as
\beq
\label{f:ATh}
\DD_{\AA} \mathbf{\Th}_{ab} = \PP^{-1} \mathbf{F}(\{ \PP \mathbf{\Th}_{ab} \} )
\eeq
Rearranging the entries in a proper way, $\AA$ can be written as a block matrix of $2\times 2$ elements per block, and each block can be diagonalized separately, with eigenvalues
\beq
\l^{(k\pm )}=1 \pm \b^2 \sqrt{f(k)f(p-k)}
\label{f:eigen}
\eeq
and eigenvectors:
\beq
v^{k\pm}= \begin{bmatrix}  \frac{1}{2 \sqrt{f(p-k)}} \\ \mp  \frac{1}{2 \sqrt{f(k)}}\end{bmatrix} \ .
\eeq
For each block of the matrix, labeled by $k$, the eigenvector matrix and its inverse,  thus, are
\beq
\PP= \begin{bmatrix} \frac{1}{2 \sqrt{f(p-k)}} &  \frac{1}{2 \sqrt{f(p-k)}} \\  -\frac{1}{2 \sqrt{f(k)}} &   \frac{1}{2 \sqrt{f(k)}} \end{bmatrix}
\eeq
\beq
\PP^{-1}= \begin{bmatrix}  \sqrt{f(p-k)} &  -\sqrt{f(k)} \\   \sqrt{f(p-k)} &   \sqrt{f(k)} \end{bmatrix}
\eeq
with 
\beq
f(k)=\frac{1}{M^{2k-1}}\binom{M}{k}
\eeq
and 
  \begin{eqnarray}
 \Th^{(k+)}&=&\sqrt{f(p-k)} Q^{(k)} - \sqrt{f(k)} Q^{(p-k)}\quad
\label{f:Thkp}
\\
 \Th^{(k-)}&=&\sqrt{f(p-k)} Q^{(k)} + \sqrt{f(k)} Q^{(p-k)}\quad
 \label{f:Thkm}
 \end{eqnarray}
In the present case, since $p$ is odd, the only overlaps we need to consider are
 $Q^{(\frac{p-1}{2})}$ and $Q^{(\frac{p+1}{2})}$. Their self-consistency equation can
 be written in the form:
 \beq
 \AA \begin{bmatrix} Q^{(\frac{p-1}{2})}_{ab} \\Q^{(\frac{p+1}{2})}_{ab} \end{bmatrix} =\begin{bmatrix}F_{\frac{p-1}{2}} (\mathbf{Q})  \\ F_{\frac{p+1}{2}} (\mathbf{Q})\end{bmatrix}
 \label{f:diag2}
 \eeq 
with
\beq
\AA= \begin{bmatrix} 1 &   - \b^2 f(\frac{p-1}{2}) \\   - \b^2 f(\frac{p+1}{2})  &   1 \end{bmatrix} \ .
\label{f:A2}
\eeq 
 The functions  $F_{(p\pm 1)/2}$ are two polinomials of degree two in all
 the $Q^{(k)}$'s.  However, as mentioned above, in order to study the
 nature of the critical behavior (continuous or discontinuous) we only
 need the terms relevant at the highest critical temperature,
 cf. Eq. (\ref{f:betac}), and we, thus, set to zero all the overlap
 matrices except for $Q^{(\frac{p-1}{2})}$ and $Q^{(\frac{p+1}{2})}$.
%
\\
\indent
Depending on the parity of $(p+1)/2$ the relevant
terms contributing to the nonlinear couplings in the action
Eq. (\ref{f:actionQ3}) differ. We now consider the two cases separately.
 \subsubsection{$(p+1)/2$ even}
 If $p=4a-1$ with $a\in \mathbb{N}$
 the functions on the r.h.s. of Eq. (\ref{f:diag2}) read:
\begin{eqnarray}
&&F_{\frac{p-1}{2}}(\mathbf{Q})= \frac{\b^4}{M^{\frac{3}{2} p - \frac{7}{2}}} \binom{M}{\frac{p-1}{2}} 
\sum_{c=1}^n Q_{ac}^{(\frac{p+1}{2})} Q_{cb}^{(\frac{p+1}{2})}  
\nonumber
\\
\nonumber
&&+ \frac{\b^4}{M^{\frac{3}{2} p - \frac{5}{2}}} 
 \binom{M}{\frac{p-1}{2}} \binom{\frac{p-1}{2}}{\frac{p+1}{4}} \binom{M-\frac{p-1}{2}}{\frac{p+1}{4}}
Q_{ab}^{(\frac{p-1}{2})} Q_{ab}^{(\frac{p+1}{2})}
\\
\end{eqnarray}

\begin{eqnarray}
&&F_{\frac{p+1}{2}}(\mathbf{Q})  =
 \frac{\b^4}{M^{\frac{3}{2} p -
\frac{1}{2}}} \binom{M}{\frac{p+1}{2}} \sum_{c=1}^n Q_{ac}^{(\frac{p-1}{2})}
Q_{cb}^{(\frac{p-1}{2})} 
\nonumber
\\
\nonumber
&&+ \frac{\b^4}{2M^{\frac{3}{2} p - \frac{1}{2}}}
\binom{M}{\frac{p+1}{2}}
\binom{\frac{p+1}{2}}{\frac{p+1}{4}}
\binom{M-\frac{p+1}{2}}{\frac{p+1}{4}}
\left(Q_{ab}^{(\frac{p-1}{2})}\right)^2
\\
\nonumber
&&+\frac{\b^4}{2M^{\frac{3}{2} p
- \frac{5}{2}}} 
 \binom{M}{\frac{p+1}{2}}
\binom{\frac{p+1}{2}}{\frac{p+1}{4}} 
\binom{M-\frac{p+1}{2}}{\frac{p-3}{4}}
\left(Q_{ab}^{(\frac{p+1}{2})}\right)^2
\\
\end{eqnarray}

\subsubsection{$(p-1)/2$ even}
If, otherwise, $p=4a+1$ with $a\in \mathbb{N}$, one obtains
\begin{eqnarray}
&&F_{\frac{p-1}{2}}  = \frac{\b^4}{M^{\frac{3}{2} p - \frac{7}{2}}} 
\binom{M}{\frac{p-1}{2}} \sum_c Q_{ac}^{(\frac{p+1}{2})} Q_{cb}^{(\frac{p+1}{2})}  
\nonumber
\\
\nonumber
&&+
\frac{\b^4}{2M^{\frac{3}{2} p - \frac{7}{2}}} 
\binom{M}{\frac{p-1}{2}} \binom{\frac{p-1}{2}}{\frac{p-1}{4}}
 \binom{M-\frac{p-1}{2}}{\frac{p-1}{4}} 
\left(Q_{ab}^{(\frac{p+1}{2})}\right)^2
\\
\nonumber
 &&+ \frac{\b^4}{2M^{\frac{3}{2} p - \frac{3}{2}}} 
\binom{M}{\frac{p-1}{2}} \binom{\frac{p-1}{2}}{\frac{p-1}{4}}
   \binom{M-\frac{p-1}{2}}{\frac{p+3}{4}}
\left(Q_{ab}^{(\frac{p-1}{2})}\right)^2
\\
\end{eqnarray}
\begin{eqnarray}
&&F_{\frac{p+1}{2}}= \frac{\b^4}{M^{\frac{3}{2} p - \frac{1}{2}}} 
\binom{M}{\frac{p+1}{2}} \sum_c Q_{ac}^{(\frac{p-1}{2})} Q_{cb}^{(\frac{p-1}{2})} 
\nonumber
\\
\nonumber
&&+ \frac{\b^4}{M^{\frac{3}{2} p - \frac{3}{2}}} 
 \binom{M}{\frac{p+1}{2}} \binom{\frac{p+1}{2}}{\frac{p-1}{4}} \binom{M-\frac{p+1}{2}}{\frac{p-1}{4}}
Q_{ab}^{(\frac{p-1}{2})} Q_{ab}^{(\frac{p+1}{2})}
\\
\end{eqnarray}

\subsection*{Computation of the coupling constants}
In order to decouple Eqs. (\ref{f:diag2})-(\ref{f:A2}) we specify the two new variables, Eqs. (\ref{f:Thkp})-(\ref{f:Thkm}), for $k=(p-1)/2$:
\beq \Th_{ab}^{(+)}=\sqrt{f\left( \frac{p+1}{2}\right)}
 Q_{ab}^{(\frac{p-1}{2})} - \sqrt{f\left( \frac{p-1}{2}\right)}
 Q_{ab}^{(\frac{p+1}{2})} \eeq 
\beq \Th_{ab}^{(-)}=\sqrt{f\left(
 \frac{p+1}{2}\right)} Q_{ab}^{(\frac{p-1}{2})} + \sqrt{f\left(
 \frac{p-1}{2}\right)} Q_{ab}^{(\frac{p+1}{2})} \eeq 
Applying the diagonalization transformation described above,
 cf. Eqs. (\ref{f:AQ})-(\ref{f:ATh}), one finds
\begin{eqnarray}
  \l^{(+)} \Th^{(+)}_{ab}&=& \sqrt{f\left( \frac{p+1}{2}\right)} F_{\frac{p-1}{2}}  - \sqrt{f\left( \frac{p-1}{2}\right)}F_{\frac{p+1}{2}} 
\nonumber
\\
  \label{f:eigen2a}
\\
  \l^{(-)} \Th^{(-)}_{ab}&=& \sqrt{f\left( \frac{p+1}{2}\right)} F_{\frac{p-1}{2}}  + \sqrt{f\left( \frac{p-1}{2}\right)} F_{\frac{p+1}{2}}
\nonumber
\\ \label{f:eigen2}
 \end{eqnarray}
where the eigenvalues, cf.  Eq. (\ref{f:eigen}), are
\begin{equation}
\l^{(\pm)} = 1\pm\beta^2\sqrt{f((p-1)/2)
 f((p+1)/2)}
\nonumber
\end{equation}
  Since the $F$'s depend on
 the $Q$'s, we have to apply the inverse transformation to get
 equations in terms of the $\Th$'s.  The eigenvalue $\l^{(+)}$ is always
 positive, so that $\Th^{(+)}$ plays the same role of the ``non
 critical'' overlaps and can be put to zero. The inverse
 transformation, thus, reduces to 
\beq
 \begin{split}
 & Q_{ab}^{(\frac{p-1}{2})}=\frac{\Th_{ab}^{(-)}}{2\sqrt{f\left( \frac{p+1}{2}\right)}} \\
 & Q_{ab}^{(\frac{p+1}{2})}=\frac{\Th_{ab}^{(-)}}{2\sqrt{f\left( \frac{p-1}{2}\right)}}
 \end{split}
 \eeq
 so that Eq. (\ref{f:eigen2}) decouples in
 \begin{eqnarray}
 \l^{(-)} \Th^{(-)}_{ab}&=& w_1(p,M)\,\, \sum_c \Th^{(-)}_{ac} \Th^{(-)}_{cb} 
 \nonumber
 \\
 &&\quad + w_2(p,M)\,\, (\Th^{(-)}_{ab})^2 \ .
 \end{eqnarray}
 The constants $w_1$ and $w_2$ depend on $p$ and $M$. The expression for $w_1$ is 
 \beq
  w_1(p,M)=\frac{1}{4M^{p-3/2}}\left(
  \sqrt{\binom{M}{\frac{p-1}{2}}}+ \sqrt{\binom{M}{\frac{p+1}{2}}}
  \right)
 \eeq
The formula for $w_2$ changes depending on the parity of $(p+1)/2$.
For even $(p+1)/2$:
   \begin{eqnarray}
    w_2(p,M) &=&\frac{1}{8M^{p-3/2}} 
     \sqrt{  \binom{M}{\frac{p-1}{2}}}
\\
\nonumber
&&
\times         \Biggl[
        \binom{\frac{p-1}{2}}{\frac{p+1}{4}}
   \binom{M-\frac{p-1}{2}}{\frac{p+1}{4}} 
 \\
 \nonumber
 &&
\quad  +         \binom{\frac{p+1}{2}}{\frac{p+1}{4}} 
        \binom{M-\frac{p+1}{2}}{\frac{p+1}{4}}
    \\
 \nonumber
 && 
 \quad+
\frac{2M-p+1}{1+p}      \binom{\frac{p+1}{2}}{\frac{p+1}{4}} 
 \binom{M-\frac{p+1}{2}} {\frac{p-3}{4}} \Biggr]
\end{eqnarray}
If $(p+1)/2$ is odd it reads:
 \begin{eqnarray}
 w_2(p,M) &=& \frac{1}{8 M^{p-\frac{3}{2}}}\sqrt{ \binom{M}{\frac{p+1}{2}}}
 \\
 \nonumber
&& \times
 \Biggl[
    \binom{\frac{p-1}{2}}{\frac{p-1}{4}}  
         \binom{M-\frac{p-1}{2}}{\frac{p-1}{4}} 
     \\
     \nonumber
&&   \quad  +  \frac{1+p}{2M-p+1}  \binom{\frac{p-1}{2}}{\frac{p-1}{4}} 
     \binom{M-\frac{p-1}{2}} {\frac{p+3}{4}}  
     \\
     \nonumber
     &&\quad
     +  2
  \binom{\frac{p+1}{2}}{\frac{p-1}{4}} \binom{M-\frac{p+1}{2}}{\frac{p-1}{4}} 
\Biggr]
\end{eqnarray}

\end{document}